\documentclass[twocolumn,letter]{jpsj3}
\def\runauthor{D. Aoki \textit{et al.}}

\title{Large Upper Critical Field of Superconductivity in the Single Crystal U$_6$Co}

\author{
Dai~Aoki$^{1,2}$\thanks{E-mail: aoki@imr.tohoku.ac.jp}, 
Ai~Nakamura$^1$,
Fuminori~Honda$^1$,
DeXin~Li$^1$, and
Yoshiya~Homma$^1$
}

\inst{%
$^1$IMR, Tohoku University, Oarai, Ibaraki, 311-1313, Japan\\
$^2$INAC/PHELIQS, CEA-Grenoble, 38054 Grenoble, France
}

\abst{%
We grew single crystals of U$_6$Co by the self-flux method and measured the magnetic susceptibility, resistivity, and specific heat.
The magnetic susceptibility shows very small anisotropy and weak temperature dependence, indicating small spin-susceptibility.
Superconductivity was clearly observed in the resistivity, susceptibility, and specific heat at $T_{\rm c}\sim 2.3\,{\rm K}$.
The upper critical field was remarkably large, $7.9$ and $6.6\,{\rm T}$ for $H\parallel [001]$ and $[110]$, respectively, in the tetragonal structure,
indicating that the ellipsoidal Fermi surface is slightly suppressed along the $[001]$ direction according to the effective mass model.
The specific heat shows a large jump at $T_{\rm c}$ with $\Delta C/\gamma T_{\rm c}=1.58$,
and the field dependence of the specific heat at low temperatures shows an almost linear increase.
These experimental results are well explained by the BCS model in the dirty limit condition. 
U$_6$Co is most likely a conventional $s$-wave superconductor with a full superconducting gap.
}

\begin{document}
\maketitle
%------------------------------
Superconductivity is one of the most fascinating topics in condensed matter physics. 
In particular, there are many unconventional superconductors correlated with magnetism in heavy fermion systems,
A simple criterion for the conventional superconductivity is the so-called Hill limit~\cite{Hil70},
which indicates the distance between $f$-electron atoms and the direct overlap of the $f$-wave functions.
If the distance is lower than the Hill limit ($d<3.5\,{\rm \AA}$ for uranium compounds),
the wave function of $f$-electrons forms a narrow band of itinerant electrons and the high density of states at the Fermi level,
which is favorable for conventional superconductivity with phonon interactions.
On the other hand, a large distance with $d > 3.5\,{\rm \AA}$ creates a magnetic order, 
which is unfavorable for conventional superconductivity.

U$_6$X (X=Mn, Fe, Co, Ni) is an interesting system that shows superconductivity with a relatively high $T_{\rm c}$ compared with other uranium-based superconductors.~\cite{Cha58,Hil68}
The values of $T_{\rm c}$ of these materials are 
$2.31\,{\rm K}$ (U$_6$Mn),
$3.78\,{\rm K}$ (U$_6$Fe),
$2.33\,{\rm K}$ (U$_6$Co), and
$0.33\,{\rm K}$ (U$_6$Ni).
Indeed, U$_6$Co has the second highest $T_{\rm c}$ among all the known uranium-based superconductors.
By alloying the compounds from Mn to Ni, which tunes the number of valence electrons of the transition-element,
$T_{\rm c}$ changes smoothly and takes a maximum at U$_6$Fe.
This tendency had been discussed from the viewpoint of the correlation between magnetism and superconductivity
because of the analogy of Slater--Pauling curves.~\cite{Hil68}
Note that U$_6$X was suggested to be close to the boundary between superconducting and ferromagnetic ground states,~\cite{Del85}
which is now extensively studied for ferromagnetic superconductors in heavy fermion systems~\cite{Aok12_JPSJ_review}.
The smooth variation of $T_{\rm c}$ also indicates that the superconductivity in this system is robust
against disorder and impurities,
suggesting conventional superconductivity.
The upper critical field $H_{\rm c2}$ for $T\to 0$ is, however, markedly large in U$_6$Fe and U$_6$Co.
In U$_6$Fe, for example, $H_{\rm c2}(0)$ is $13.1$ and $10.4\,{\rm T}$ for $H \parallel [001]$ and $[110]$, respectively, in the tetragonal structure.~\cite{Yam96}.
There are no reports on $H_{\rm c2}(0)$ in U$_6$Co down to low temperatures, 
but a large $H_{\rm c2}(0)$ is inferred from the large initial slope of $H_{\rm c2}$.~\cite{Del87}
The penetration depth~\cite{Hou87} and NMR~\cite{Koh87} experiments suggest BCS-type conventional superconductivity in U$_6$Co.
All the previous experiments in U$_6$Co were done with polycrystalline samples.
Although conventional superconductivity is most likely realized in U$_6$Co,
it is important to clarify the superconducting properties using a single crystal, especially focusing on the large $H_{\rm c2}$.
Therefore, we grew single crystals of U$_6$Co and measured the resistivity, magnetic susceptibility, and specific heat down to low temperatures.

\begin{figure}[tbh]
\begin{center}
\includegraphics[width=1 \hsize,clip]{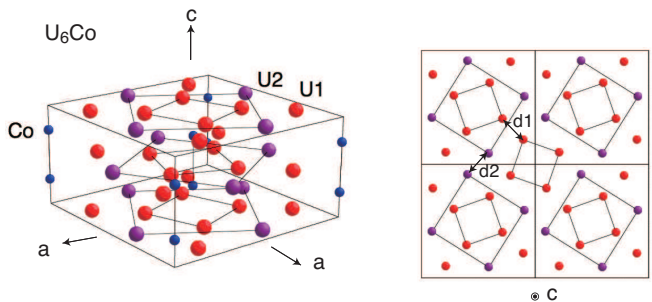}
\end{center}
\caption{(Color online) Tetragonal crystal structure of U$_6$Co. The U atom has two sites as denoted by U1 and U2. The distances for the first and second nearest neighbors of the U atom are denoted by d1 and d2, respectively.}
\label{fig:U6Co_structure}
\end{figure}

U$_6$Co forms the U$_6$Mn-type tetragonal crystal structure with the space group $I4/mcm$ (\#140, $D_{4h}^{18}$), as shown in Fig.~\ref{fig:U6Co_structure}.
The unit cell containing four molecules is relatively large with the lattice parameters of $a=10.36\,{\rm \AA}$ and $c=5.21\,{\rm \AA}$.
The two sites of the U atom, U1 and U2 are located in the $c$-plane, forming square lattices for each U site.
These U-planes are stacked along the $c$-axis, and the square lattice in the adjacent U-plane is slightly rotated in the $c$-plane.
The Co atom is located at $z=0.25$ between the U-planes.
The distance between U atoms is quite small.
The first nearest neighbor is $d1=2.68\,{\rm \AA}$ between U1 sites, and 
the second nearest neighbor is $d2=2.72\,{\rm \AA}$ between U2 sites.
These values are far below the Hill limit,
implying the direct overlap of 5$f$-wave functions and the itinerant nature of 5$f$-electrons.
%========================================================================================
%========================================================================================

\begin{figure}[tbh]
\begin{center}
\includegraphics[width=1 \hsize,clip]{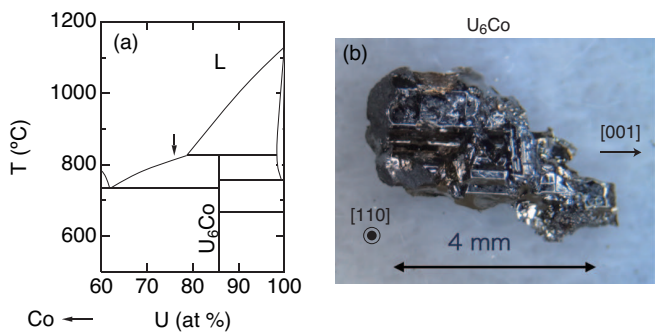}
\end{center}
\caption{(Color online) (a) U-Co alloy phase diagram. The data are taken from Ref.~\protect\citen{Mas90}. The arrow indicates the starting composition for the self-flux method. (b) Photograph of a U$_6$Co single crystal grown by the self-flux method.}
\label{fig:U6Co_photo}
\end{figure}

Single crystals of U$_6$Co were grown by the self-flux method.
Figure~\ref{fig:U6Co_photo}(a) shows the binary phase diagram of U-Co~\cite{Mas90}.
The conventional Czochralski pulling method with stoichiometric amounts is not applicable.
Thus, we attempted to grow U$_6$Co single crystals by the self-flux method.
The starting materials of U and Co with off-stoichiometric amounts,
U:Co = 76:24 (atomic {\%}),
 shown in Fig.~\ref{fig:U6Co_photo}(a), were put in an alumina crucible,
which was sealed in a quartz ampoule.
The quartz ampoule was then heated to $830\,^\circ{\rm C}$ in an electrical furnace.
The temperature was maintained for $300\,{\rm h}$ and
then decreased to $740\,^\circ{\rm C}$ with a slow cooling rate of $0.4\,^{\circ}{\rm C/h}$.
The excess Co was removed by spinning off in the centrifuge at this temperature.
Many single crystals of U$_6$Co with a bar shape elongated along the $c$-axis were obtained,
as shown in Fig.~\ref{fig:U6Co_photo}(b).
The single crystals were oriented using the Laue photograph, and were cut with a spark cutter.
%========================================================================================
%========================================================================================

The magnetic susceptibility was measured using a commercial SQUID magnetometer 
at temperatures down to $2\,{\rm K}$.
The specific heat was measured at temperatures down to $0.36\,{\rm K}$ and at fields up to $9\,{\rm T}$
using a relaxation technique in a $^3$He cryostat.
The resistivity was measured by the four-probe AC method 
at temperatures down to $0.03\,{\rm K}$ and at fields up $14\,{\rm T}$
in a $^3$He cryostat and in a dilution refrigerator.

\begin{figure}[tbh]
\begin{center}
\includegraphics[width=0.8 \hsize,clip]{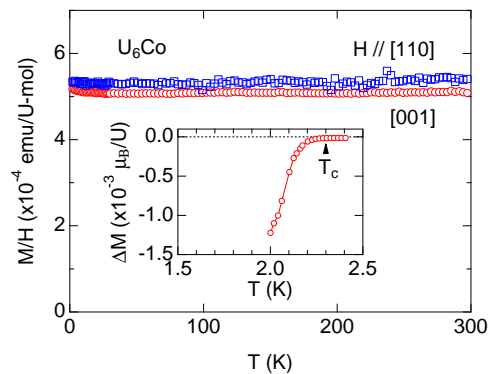}
\end{center}
\caption{(Color online) Temperature dependence of the magnetic susceptibility at $H=1\,{\rm T}$ for $H\parallel [110]$ and $[001]$ in U$_6$Co. The inset shows the temperature dependence of magnetization at low temperatures measured below $0.001\,{\rm T}$.}
\label{fig:U6Co_sus}
\end{figure}

%~~~~~~~~~~~~~~~~~~~~~~~~~~~~~~~~~~~~~~~~~~~~~~~~~~~~~~~~~~~~~~~~~~~~~~~~~~~~~~~~~~~~~~~~~~~~~~~~~~~~~~~~~~~~~~~~~~~~~~~~~~~~~
Figure~\ref{fig:U6Co_sus} shows the temperature dependence of magnetic susceptibility 
for $H\parallel [001]$ and $[110]$.
The susceptibility shows quite small anisotropy and
almost temperature-independent behavior down to $2.3\,{\rm K}$,
indicating Pauli paramagnetism.
Below $2.3\,{\rm K}$, the susceptibility starts to decrease,
revealing a diamagnetic signal due to the superconducting transition,
as shown in the inset of Fig.~\ref{fig:U6Co_sus}.
%========================================================================================
%========================================================================================

\begin{figure}[tbh]
\begin{center}
\includegraphics[width=0.8 \hsize,clip]{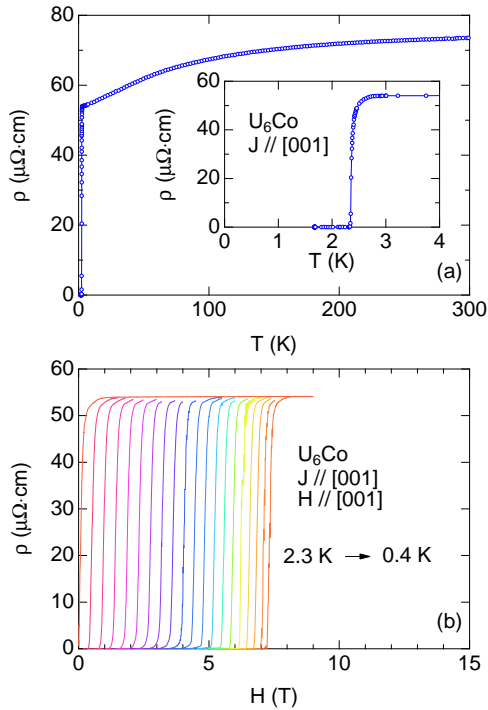}
\end{center}
\caption{(Color online) (a) Temperature dependence of the resistivity for $J \parallel [001]$ in U$_6$Co. The inset shows the low-temperature part. (b) Field dependence of the longitudinal magnetoresistance for $J \parallel [001]$ and $H\parallel [001]$ at different constant temperatures from $2.3$ to $0.4\,{\rm K}$ with $0.1\,{\rm K}$ steps.}
\label{fig:U6Co_resist}
\end{figure}

Figure~\ref{fig:U6Co_resist}(a) shows the temperature dependence of resistivity for the electrical current along the $[001]$ direction.
The resistivity decreases with decreasing temperature with a convex curvature
and shows a superconducting transition.
The resistivity starts to drop around $2.8\,{\rm K}$ and becomes zero at $2.33\,{\rm K}$.
The midpoint of the resistivity drop, which we define as the superconducting transition temperature hereafter, 
is $T_{\rm c}=2.36\,{\rm K}$.
The resistivity follows $T^2$ dependence below $7\,{\rm K}$, that is, $\rho=\rho_0+AT^2$,
with the coefficient $A=0.012\,\mu\Omega\!\cdot\!{\rm cm\,K^{-2}}$ and 
the residual resistivity $\rho_0=54\,\mu\Omega\!\cdot\!{\rm cm}$.
Despite the large residual resistivity, the superconducting transition is rather sharp,
indicating that the superconductivity of U$_6$Co is not very sensitive to the sample quality.
This also implies conventional superconductivity in U$_6$Co.
%========================================================================================
%========================================================================================
 
\begin{figure}[tbh]
\begin{center}
\includegraphics[width=0.8 \hsize,clip]{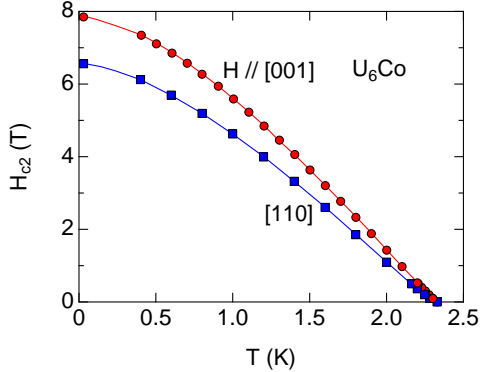}
\end{center}
\caption{(Color online) Temperature dependence of the upper critical field $H_{\rm c2}$ for $H\parallel [001]$ and $[110]$ determined by the field and temperature dependences of resistivity in U$_6$Co. The solid lines are guides to the eyes.}
\label{fig:U6Co_HT_phase}
\end{figure}

In order to determine the superconducting phase diagram, we measured the magnetoresistance for $H\parallel [001]$ and $[110]$. 
Figure~\ref{fig:U6Co_resist}(b) shows the magnetoresistance at different temperatures for $H\parallel [001]$.
The superconducting transition is sharp even at high fields, thus, we could determine the temperature dependence of the upper critical field $H_{\rm c2}$, as shown in Fig.~\ref{fig:U6Co_HT_phase}.
Here, we define $H_{\rm c2}$ at the midpoint of the resistivity drop.
The values of $H_{\rm c2}$ are remarkably large for both $H\parallel [001]$ and $[110]$.
At the lowest temperature, $0.03\,{\rm K}$, $H_{\rm c2}$ is $7.85\,{\rm T}$ for $H\parallel [001]$
and $6.56\,{\rm T}$ for $H\parallel [110]$.
The initial slope of $H_{\rm c2}$ is $-dH_{\rm c2}/dT = 4.3\,{\rm T/K}$ for $H \parallel [001]$ and $3.4\,{\rm T/K}$ for $H\parallel [110]$.
In general, the Pauli limiting field is given by $H_{\rm P}=\sqrt{2}\Delta/(g\mu_{\rm B}) = 1.86 T_{\rm c}$,
assuming the $g$-factor $g=2$ and the superconducting gap $\Delta$, with $2\Delta /(k_{\rm B}T_{\rm c})=3.53$ based on the weak coupling BCS model.
In this case, $H_{\rm P}$ in U$_6$Co is estimated to be $4.3\,{\rm T}$, which is much smaller than the
measured $H_{\rm c2}$ for both directions.
Thus, strong-coupling superconductivity or a small $g$-factor is required if we assume the spin-singlet state.
As shown later, the specific heat jump at $T_{\rm c}$, namely, $\Delta C/\gamma T_{\rm c}$ is
only $10{\%}$ larger than that for the weak coupling BCS model.
Therefore, a small $g$-factor is expected.
In fact, the weakly temperature dependent susceptibility for $[001]$ and $[110]$
indicates very small spin-susceptibility, which is consistent with a small $g$-value.
For $H_{\rm c2} \leq H_{\rm P}$, one can expect $g < 1.2$.
%========================================================================================
%========================================================================================

\begin{figure}[tbh]
\begin{center}
\includegraphics[width=0.8 \hsize,clip]{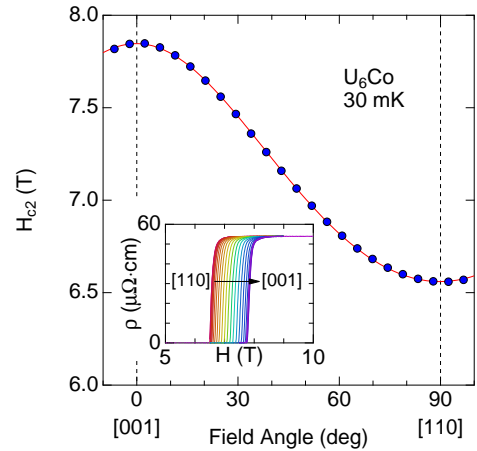}
\end{center}
\caption{(Color online) Angular dependence of the upper critical field $H_{\rm c2}$ at $30\,{\rm mK}$ from $H\parallel [001]$ to $[110]$. The solid line shows the results of fitting by the effective mass model. The inset shows the field dependence of magnetoresistance for different field directions. $H_{\rm c2}$ was defined as the midpoint of the resistivity drop.}
\label{fig:U6Co_Hc2_AngDep}
\end{figure}

We show in Fig.~\ref{fig:U6Co_Hc2_AngDep} the angular dependence of $H_{\rm c2}$ determined by the magnetoresistance at $30\,{\rm mK}$.
By rotating the field angle, $H_{\rm c2}$ smoothly increases from $6.56\,{\rm T}$ for $[110]$ to $7.85\,{\rm T}$
for $[001]$.
The angular dependence of $H_{\rm c2}$ can be fitted by the so-called effective mass model,
where an ellipsoidal Fermi surface with anisotropic effective mass is assumed.
The solid line in the main panel of Fig.~\ref{fig:U6Co_Hc2_AngDep} is the result of fitting 
by the equation 
$H_{\rm c2}(\theta )=H_{\rm c2}(90^\circ)/\sqrt{\sin^2\theta + (m_c/m_a)\cos^2\theta}$.
We obtain $m_c/m_a = 0.699$ and $k_c/k_a = 0.84$.
The effective Fermi surface is almost spherical with slight suppression along the $[001]$ direction.
%========================================================================================
%========================================================================================

\begin{figure}[tbh]
\begin{center}
\includegraphics[width=0.9 \hsize,clip]{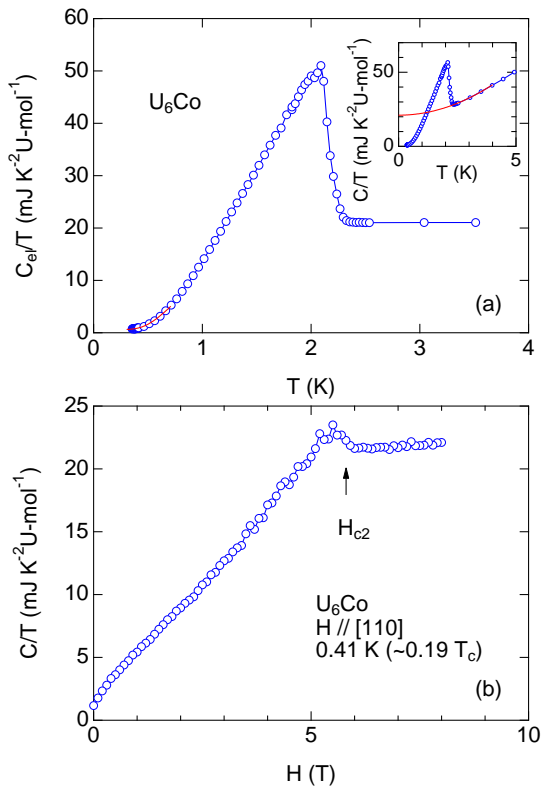}
\end{center}
\caption{(Color online) (a) Temperature dependence of the electronic specific heat at zero field in the form of $C_{\rm el}/T$ vs $T$ in U$_6$Co.
The red solid line below $0.71\,{\rm K}$ shows the results of fitting by the BCS full gap model.
 The inset shows the total specific heat in the form of $C/T$ vs $T$.
The red solid line in the inset shows the specific heat in the normal state extrapolated to $0\,{\rm K}$
as a function of $C/T = \gamma + \beta T^2$.
(b) Field dependence of the specific heat in the form of $C/T$ vs $H$ for $H\parallel [110]$ at $0.41\,{\rm K}$, which is approximately equal to $0.19\,T_{\rm c}$.
}
\label{fig:U6Co_Cp}
\end{figure}

Figure~\ref{fig:U6Co_Cp}(a) shows the temperature dependence of the specific heat.
The inset of Fig.~\ref{fig:U6Co_Cp}(a) shows the total specific heat in the form of $C/T$ vs $T$.
The specific heat jump indicates bulk superconductivity.
In the normal state below $4\,{\rm K}$, the data is well fitted by $C/T = \gamma + \beta T^2$.
By extrapolating the fitting to $0\,{\rm K}$, we subtract the phonon contribution and extract the electronic specific heat $C_{\rm el}$
from the total specific heat,
as shown in the main panel of Fig.~\ref{fig:U6Co_Cp},
which also satisfies the entropy balance.
The Sommerfeld coefficient is $\gamma \simeq 20\,{\rm mJ\,K^{-1}U\mbox{-}mol^{-1}}$,
indicating moderately enhanced heavy electrons.
The value of $A/\gamma^2$ is approximately consistent with the so-called Kadowaki--Woods ratio.
$T_{\rm c}$ determined by the entropy balance is $2.17\,{\rm K}$, which is slightly lower than the value of $2.33\,{\rm K}$ from the resistivity measurements, but is not very far.
The jump of the specific heat is $\Delta C/\gamma T_{\rm c}=1.58$,
which is slightly larger but close to the weak coupling BCS value of $1.43$.

The low-temperature data from $0.35$ to $0.71\,{\rm K}$ were well fitted by the BCS asymptotic formula~\cite{Kre74},
$C_{\rm el}/T = \gamma_0+3.15 (\Delta/1.76 k_{\rm B}T)^{5/2}\gamma \exp (-\Delta/k_{\rm B}T)$,
as shown in the main panel in Fig.~\ref{fig:U6Co_Cp}(a).
We obtain the gap energy, $\Delta = 3.97\,{\rm K}$, and $2\Delta / k_{\rm B}T_{\rm c}=3.7$,
which is also close to the weak coupling BCS value.

The thermodynamic critical field $H_{\rm c}$ can be calculated from
the difference in free energy between the superconducting and normal states,
that is,
$\Delta F(T) = \mu_0 H_{\rm c}^2(T)/2 = \int_{T_{\rm c}}^T [S_{\rm s}(T)-S_{\rm n}(T)] dT$,
where $S_{\rm s}$ and $S_{\rm n}$ are the entropies in the superconducting and normal states, respectively,
which are calculated as $S=\int C_{\rm el}/T dT$.
We obtain $H_{\rm c}(0)$ as $0.065\,{\rm T}$, which roughly agrees with the value reported previously, $0.072\,{\rm T}$~\cite{Yan89}.

In order to investigate the superconducting gap structure, we measured the field dependence of the specific heat
at a low temperature, $0.41\,{\rm K}$, as shown in Fig.~\ref{fig:U6Co_Cp}(b).
The temperature is approximately equal to $0.19 T_{\rm c}$.
$C/T$ increases almost linearly with a slight convex curvature, 
implying full-gap superconductivity at the first approximation,
although measurements at lower temperatures and in different field directions are required to obtain a conclusion.
%========================================================================================
%========================================================================================

On the basis of the BCS theory assuming a spherical Fermi surface,
we calculate some parameters of the superconductivity.~\cite{Rau87}
From the starting parameters of $\gamma = 20\,{\rm mJ K^{-2}U\mbox{-}mol^{-1}}$, $\rho_0=54\,\mu\Omega\!\cdot\!{\rm cm}$, $T_{\rm c}=2.3\,{\rm K}$, and the lattice parameters,
we obtain the initial slope of $H_{\rm c2}$ as $-dH_{\rm c2}/dT=3.5\,{\rm T/K}$ in the case of the dirty limit.
This is close to the values obtained in the experiments, namely,
$3.36$ and $4.30\,{\rm T/K}$ for $H\parallel [110]$ and $[001]$, respectively.
Taking the average of the experimental initial slope, we obtain the effective Fermi surface as 
$1.6\times 10^{21}\,{\rm m^{-2}}$, ($4.3\times 10^{20}\,{\rm m^{-2}}$) for the dirty (clean) limit.
The mean free path is $l = 67\,{\rm \AA}$, ($18\,{\rm \AA}$).
The BCS coherence length for $T\to 0$ is $\xi_0= 0.18 \hbar v_{\rm F}/k_{\rm B}T_{\rm c}=84\,{\rm \AA}$, ($38\,{\rm \AA}$).
The mean free path is smaller than the coherence length, revealing that
the present sample is in the dirty limit condition.
This suggests that U$_6$Co is a conventional BCS superconductor.
The coherence lengths estimated from $H_{\rm c2}=\phi_0/2\pi\xi^2$ are
$\xi=65$ and $71\,{\rm \AA}$ for $H \parallel [001]$ and $[110]$, respectively.
These values are close to the BCS coherence length $\xi_0$ estimated in the dirty limit.

Note that similar $H_{\rm c2}$ curves have also been reported for U$_6$Fe.~\cite{Yam96}
Since U$_6$Fe has a slightly larger $T_{\rm c}$ ($\sim 4\,{\rm K}$) than U$_6$Co,
$H_{\rm c2}$ is also large, $13.1\,{\rm T}$ for $H\parallel [001]$ and $10.4\,{\rm T}$ for $[110]$.
These values can be scaled by a factor of $1.7$ from those of U$_6$Co.
The angular dependence of $H_{\rm c2}$ with the effective mass anisotropy $m_c/m_a=0.64 $ in U$_6$Fe 
is also similar to that of U$_6$Co.

%-----------------
In summary, we succeeded in growing single crystals of U$_6$Co by the self-flux method.
The magnetic susceptibility is almost temperature-independent and is quite isotropic between $H\parallel [001]$ and $[110]$,
indicating small spin-susceptibility compared with orbital susceptibility,
which is consistent with the small $g$-factor estimated from the specific heat and the Pauli limit.
The upper critical field $H_{\rm c2}$ and its initial slope are relatively large,
indicating a moderately enhanced heavy fermion system.
The anisotropy of $H_{\rm c2}$ can be well explained by the effective mass model, 
which reveals the almost spherical Fermi surface suppressed slightly along the $[001]$ direction.
The superconducting parameters obtained using the BCS model in the dirty limit condition agree reasonably well with the experiments,
indicating an $s$-wave superconductor with most likely a full gap of U$_6$Co.

\section*{Acknowledgements}
We thank J. P. Brison, Y. \={O}nuki, K. Ishida, and Y. Shimizu for fruitful discussion.
This work was supported by ERC starting grant (NewHeavyFermion), KAKENHI (25247055, 15H05884, 15H05882, 15K21732, 16H04006), ICC-IMR, and REIMEI.

%\bibliographystyle{myjpsj}
%\bibliography{bibbase}

\begin{thebibliography}{10}
\expandafter\ifx\csname url\endcsname\relax
  \def\url#1{\texttt{#1}}\fi
\expandafter\ifx\csname urlprefix\endcsname\relax\def\urlprefix{URL }\fi

\bibitem{Hil70}
H.~H. Hill, {\em Plutonium and Other Actinides}, ed. W.~N. Miner (AIME, New
  York, 1970) p.~2.

\bibitem{Cha58}
B.~S. Chandrasekhar and J.~K. Hulm, J. Phys. Chem. Solids {\bf 7}, 259 (1958).

\bibitem{Hil68}
H.~H. Hill and B.~T. Matthias, Phys. Rev. {\bf 168}, 464 (1968).

\bibitem{Del85}
L.~E. {DeLong}, R.~P. Guertin, S.~Hasanain, and T.~Fariss, Phys. Rev. B {\bf
  31}, 7059 (1985).

\bibitem{Aok12_JPSJ_review}
D.~Aoki and J.~Flouquet, J. Phys. Soc. Jpn. {\bf 81}, 011003 (2012).

\bibitem{Yam96}
E.~Yamamoto, M.~Hedo, Y.~Inada, T.~Ishida, Y.~Haga, and Y.~\={O}nuki, J. Phys.
  Soc. Jpn. {\bf 65}, 1034 (1996).

\bibitem{Del87}
L.~E. DeLong, L.~N. Hall, S.~K. Malik, G.~W. Crabtree, W.~Kwok, and K.~A.
  {Gschneidner, Jr}, J. Magn. Magn. Mater. {\bf 63-64}, 478 (1987).

\bibitem{Hou87}
M.~K. Hou, C.~Y. Huang, and C.~E. Olsen, Solid State Commun. {\bf 61}, 101 (1987).

\bibitem{Koh87}
Y.~Kohori, T.~Kohara, Y.~Kitaoka, K.~Asayama, M.~B. Maple, and M.~W.
  {McElfresh}, J. Phys. Soc. Jpn. {\bf 56}, 1645 (1987).

\bibitem{Mas90}
T.~B. Massalski, H.~Okamoto, P.~R. Subramanian, and L.~Kacprzak, {\em Binary
  Alloy Phase Diagrams} (ASM International, Materials Park, 1990, 2nd ed.).

\bibitem{Kre74}
V.~Z. Kresin and V.~P. Parkhomenko, Fiz. Tverd. Tela {\bf 16}, 3363 (1974).

\bibitem{Yan89}
K.~N. Yang, M.~B. Maple, L.~E. {DeLong}, J.~G. Huber, and A.~Junod, Phys. Rev. B
  {\bf 39}, 151 (1989).

\bibitem{Rau87}
U.~Rauchschwalbe: Physica {\bf 147B}, 1 (1987).

\end{thebibliography}

\end{document}